\documentclass[aps,prl,10pt,showpacs,superscriptaddress,twocolumn,floatfix]%
             {revtex4-1}

\usepackage{graphicx,rotating}

\newcommand{\Exp}[1]{\mathrm{e}^{\mbox{\footnotesize$#1$}}}
\newcommand{\I}{\mathrm{i}}
\newcommand{\half}{\frac{1}{2}}

\newcommand{\tr}[2][]{\textnormal{tr}#1{\left\{#2\right\}}}
\newcommand{\heading}[1]{\par\textbf{#1}\quad\ignorespaces}
\newcommand{\ds}{\displaystyle}
\newcommand{\Eq}[2][Eq.~]{#1(\ref{eq:#2})}
\newcommand{\Fig}[2][Fig.~]{#1\ref{fig:#2}}
\newcommand{\column}[2][c]{{\left(\begin{array}{#1}#2\end{array}\right)}}
\newcommand{\sqcol}[2][c]{{\left[\begin{array}{#1}#2\end{array}\right]}}
\newcommand{\idlerwf}[2][\ ]{\phi^{#1}_{#2}}

\newcommand{\sds}[2][\depth]{\raisebox{0pt}[\height][#1]{\small$\ds #2$}}

\begin{document}
\title{Unambiguous path discrimination in a two-path interferometer}

\author{Yink Loong Len}\email[]{yinkloong@quantumlah.org}
\affiliation{Centre for Quantum Technologies, National University of
  Singapore, 3 Science Drive 2, Singapore 117543, Singapore} 
\affiliation{Data Storage Institute, %
             Agency for Science, Technology and Research, %
             2 Fusionopolis Way, \#08-01 Innovis, %
             Singapore 138634, Singapore}

\author{Jibo Dai}\email[]{dai\_jibo@dsi.a-star.edu.sg}
\affiliation{Centre for Quantum Technologies, National University of
  Singapore, 3 Science Drive 2, Singapore 117543, Singapore} 
\affiliation{Data Storage Institute, %
             Agency for Science, Technology and Research, %
             2 Fusionopolis Way, \#08-01 Innovis, %
             Singapore 138634, Singapore}

\author{Berthold-Georg Englert}\email[]{cqtebg@nus.edu.sg}
\affiliation{Centre for Quantum Technologies, National University of
  Singapore, 3 Science Drive 2, Singapore 117543, Singapore} 
\affiliation{Department of Physics, National University of Singapore, 2
  Science Drive 3, Singapore 117542, Singapore} 
\affiliation{MajuLab, CNRS-UNS-NUS-NTU International Joint Unit, %
  UMI 3654, Singapore} 

\author{Leonid A. Krivitsky}\email[]{Leonid-K@dsi.a-star.edu.sg}
\affiliation{Data Storage Institute, %
             Agency for Science, Technology and Research, %
             2 Fusionopolis Way, \#08-01 Innovis, %
             Singapore 138634, Singapore}

\date[]{Posted on the arXiv on 4 August 2017; %
        this second version on 17 November 2017}

\begin{abstract}
When a photon is detected after passing through an interferometer one might
wonder which path it took, and a meaningful answer can only be given if one
has the means of monitoring the photon's whereabouts.
We report the realization of a single-photon experiment for a two-path
interferometer with path marking. 
In this experiment, the path of a photon (``signal'') through a
Mach--Zehnder interferometer becomes known by unambiguous discrimination
between the two paths.
We encode the signal path in the polarization state of a
partner photon (``idler'') whose polarization is examined by a three-outcome
measurement: one outcome each for the two signal paths plus an inconclusive
outcome. 
Our results agree fully with the theoretical predictions from a
common-sense analysis of what can be said about the past of a quantum particle: 
The signals for which we get the inconclusive result have full interference
strength, as their paths through the interferometer 
\emph{cannot be known}; 
and every photon that emerges from the dark output
port of the balanced interferometer has a \emph{known path}.
\end{abstract}

\pacs{03.65.Ta, 42.50.Dv, 42.50.Xa}

\begin{widetext}
\maketitle    
\end{widetext}

\heading{Introduction}
Vaidman's three-path interferometer \cite{Vaidman:13a,Danan+3:13,Zhou+7:17}
is an asymmetric Mach--Zehnder interferometer (MZI) with a symmetric MZI
inserted into one arm; 
weak interactions are used to mark the path of that particle through the
interferometer. 
The emphasis is on photons detected at a particular one of the three exit
ports, and Vaidman describes the pre- and post-selected ensemble by the
two-state vector formalism that was introduced in Ref.~\cite{Aharonov+1:90}.
To that he adds the interpretational rule that ``the particle was in paths of
the interferometer in which there is an overlap of the forward and backward
evolving wave functions'' \cite{Vaidman:16b}, and then infers that each
particle was inside the internal loop at intermediate times without, however,
entering or leaving.
In this narrative the particle was simultaneously at several places at the
same time.
While this account of the particle's whereabouts is at odds with common sense,
and also with more traditional ways of speaking about the past of a quantum
particle, it is claimed that the experimental data reported in
Refs.~\cite{Danan+3:13} and \cite{Zhou+7:17} support Vaidman's narrative.
This triggered a lively debate about these matters; see, for example,
Refs.[4--33] in \cite{Englert+4:17}. 

As we argued recently, the experimental data do not provide evidence in
support of Vaidman's interpretational rule~\cite{Englert+4:17}. 
Our conclusion follows from a careful examination of the path-information that
can be extracted from the weak traces left by the particle when passing
through the interferometer.
When one acquires path knowledge by a fitting path-discriminating
measurement, all particles in the pre- and post-selected ensemble have a known
path through the interferometer --- \emph{one} known path for each particle
--- in full accordance with what common sense tells us.

An important ingredient in our analysis is a remarkable property of the
inner balanced internal MZI:
All particles that emerge from the dark output port --- where they can only be
found because the weak path marking slightly disturbs the balance of the MZI
--- have a knowable path through the MZI, despite the destructive interference
that completely prevents particles from exiting at this output port if the
balance is not upset.
In the larger context of the three-path interferometer, then, these particles
eventually recombine incoherently --- an immediate consequence of their
known path --- with particles from the outer path at the final 2:1 beam
splitter, and each particle detected after the beam splitter has a
known path. 

This prediction can be
tested with fittingly prepared entangled photon pairs (``idler'' and
``signal'') \cite{Englert+4:17}. 
The signal is fed into the MZI and detected at one of its exits, while 
the unambiguous discrimination of two polarization states of the idler
provides information about the path taken by the signal.
We report here a realization of such an experiment.
It confirms the theoretical predictions, in particular that each signal
detected at the dark output port has a known path through the MZI.

Experiments, in which the path of a quantum particle through an interferometer
is marked by different states of internal degrees of freedom, have a long
history. 
Important examples are neutrons and their spin \cite{Summhammer+3:82}; 
atoms and their fine structure \cite{Durr+2:98}; 
or photons and their polarization \cite{Schwindt+2:99}.
In other experiments, more closely related to our work, 
the paths of an interfering  quantum particle are
correlated with the degrees of freedom of auxiliary quantum systems:
photon paths with atom states \cite{Eichmann+6:93},
electron paths with quantum-dot states \cite{Buks+4:98}, 
or photon paths with partner-photon polarization states \cite{Herzog+3:95}.

The unambiguous state discrimination (USD) of two nonorthogonal states was
first demonstrated for photon polarization \cite{Clarke+3:01}.
USD has also been realized for two states of a nuclear spin
\cite{Waldherr+5:12}, three states of a path qutrit \cite{Mohseni+2:04},
four coherent states of a light mode \cite{Becerra+2:13},
and up to 14 orbital-angular-momentum states of a photon \cite{Agnew+4:14}.

\begin{figure}
  \centering
  \includegraphics[width=240pt]{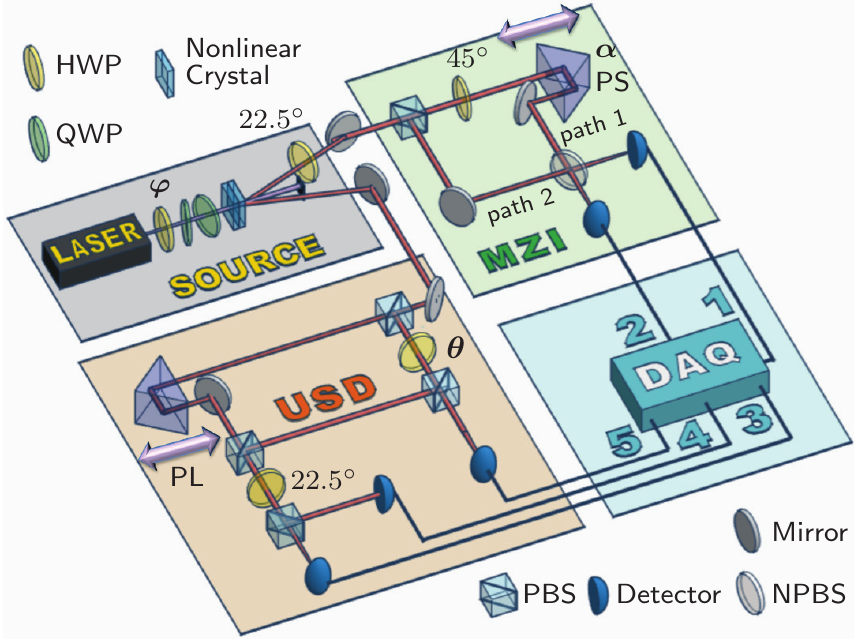}
  \caption{\label{fig:scheme}Scheme of the experiment. 
   The source emits wave\-length-selected polarization-entangled photon pairs
   (signal and idler) in the state with the wave function of \Eq{2}.
   On the signal side, a polarizing beam splitter (PBS) and a half-wave plate
   (HWP) convert the polarization qubit into the path qubit of a vertically
   polarized photon in a Mach--Zehnder interferometer (MZI).
   A phase shifter (PS) --- 
   the usual combination of a
   feedback-controlled piezo transducer and a translation stage --- 
   defines the phase $\alpha$ of the MZI.  
   When the nonpolarizing 50/50 beam splitter (NPBS) 
   is removed, the signal's detection reveals its path.
   On the idler side, the setup of Clarke \textit{et al.} \cite{Clarke+3:01}
   implements the unambiguous state discrimination (USD) of the
   two polarization states of the idler that are correlated with 
   the signal paths \cite{fn1}.
   The difference in optical length of the two paths of the loop is stabilized
   by a phase locker (PL), the analog of the PS on the signal side. 
   The photons are fed to the detectors through single-mode fibers, and 
   coincidences of any two detectors within the lapse of  \mbox{5\,ns} 
   are recorded by a data acquisition system (DAQ).
   The angle $\theta$ the HWP at the USD is set such that detection of the
   idler by detectors D3 or D4 identifies unambiguously the path followed by the
   signal; see Table~\ref{tbl:coincide} for the coincidence probabilities of
   joint detection by the idler and signal detectors.} 
\end{figure}

\heading{Scheme of the experiment}
Figure~\Fig[]{scheme} shows the scheme of the experiment.
Our photon-pair source, described in details in Ref.~\cite{Dai+4:13},
is of the kind pioneered by Kwiat \textit{et al.} \cite{Kwiat+4:99}. 
Spontaneous parametric down conversion creates photon pairs 
at 809.2\,nm from a pump laser at 404.6\,nm 
with the polarization wave function 
\begin{equation}\label{eq:1}
    \sin(2\varphi)\sqcol{1\\0}\otimes\sqcol{1\\0}
    +\cos(2\varphi)\sqcol{0\\1}\otimes\sqcol{0\\1}
\end{equation}
in $\textsc{v}/\textsc{h}$ basis,
where the first factors in the tensor products are for the idler polarization
and the second factors for the signal polarization.
The angle parameter $\varphi$ is adjusted by a half-wave plate (HWP) in the
beam of the pump laser such that both photons are
vertically polarized with probability $\sin(2\varphi)^2$ or both are
horizontally polarized with probability $\cos(2\varphi)^2$.
Before leaving the source, the signal passes through a HWP set
at $22.5^{\circ}$, which turns the wave function into
\begin{eqnarray}\label{eq:2}
   &&\sin(2\varphi)\sqcol{1\\0}\otimes\frac{1}{\sqrt{2}}\sqcol{1\\1}
     +\cos(2\varphi)\sqcol{0\\1}\otimes\frac{1}{\sqrt{2}}\sqcol{1\\-1}
\nonumber\\
   &=&\frac{1}{\sqrt{2}}\sqcol{\sin(2\varphi)\\ \cos(2\varphi)}
                 \otimes\sqcol{1\\0}
      +\frac{1}{\sqrt{2}}\sqcol{\sin(2\varphi)\\ -\cos(2\varphi)}
                 \otimes\sqcol{0\\1}\,.\nonumber\\&
\end{eqnarray}
This is the polarization-entangled idler-signal state emitted by the source.
The signal then enters the MZI, and the idler polarization is examined
by USD.

On the signal side, the combination of the polarizing beam splitter (PBS) and
the HWP at $45^\circ$ converts the polarization qubit with
the wave function $\sds{\sqcol{v'\\h'}}$ into the path qubit of a
vertically polarized photon with the wave function $\sds{\column{h'\\v'}}$
\cite{fn0}. 
The phase shifter (PS) in path~1 imprints a phase $\alpha$, so that the
probabilities of detecting the signal by detectors D1 or D2 behind the 
nonpolarizing 50/50 beam splitter (NPBS) are 
$\half\bigl|h'\Exp{\I\alpha}+v'\bigr|^2$ and
$\half\bigl|h'\Exp{\I\alpha}-v'\bigr|^2$, respectively. 
The MZI is balanced for $v'=h'$ when $\alpha=0$, with detector D2 probing the
dark output port.
See the Appendix for further details.

\begin{table}[!t]
  \centering
  \caption{\label{tbl:coincide}%
  Coincidence probabilities of detecting the signal by detector D1 or D2 and
  the idler by detector D3, D4, or D5.
  In particle mode, the NPBS is removed from the MZI in \Fig{scheme} and the PS
  is of no consequence.
  In wave mode, the NPBS is in place and the probabilities of detecting the
  signal are $2\pi$-periodic in the interferometer phase $\alpha$.}  
  \begin{tabular}{@{}lcrcccc@{}}\hline\hline
  \rule{0pt}{25pt}&\rule{5pt}{0pt}&&\rule{5pt}{0pt} 
  &\multicolumn{3}{c}{\underline{\makebox[175pt][c]{idler detector}}} \\
  MZI&& 
  \raisebox{0pt}[0pt][0pt]{\begin{rotate}{90}%
    \begin{minipage}[b]{33pt}signal detector\end{minipage}\end{rotate}} &
  & D3 & D4 & D5\\ \hline
      && D1 && $\cos(2\varphi)^2$ & 0 & 
               $-\half\cos(4\varphi)$\rule{0pt}{12pt} \\[0.7ex]
  \raisebox{3pt}[0pt][0pt]{\begin{minipage}[b]{30.7pt}\small\flushleft%
   particle mode\end{minipage}} && D2 && 0 & $\cos(2\varphi)^2$ 
   & $ -\half\cos(4\varphi)$ \\[1.5ex]
  && D1 && $\half\cos(2\varphi)^2$ & $\half\cos(2\varphi)^2$ &
         $-\cos(4\varphi)\cos\bigl(\half\alpha\bigr)^2$ \\[0.7ex]
  \raisebox{3pt}[0pt][0pt]{\begin{minipage}[b]{30pt}\small\flushleft%
   wave mode\end{minipage}}  && D2 && 
         $\half\cos(2\varphi)^2$ & $\ \half\cos(2\varphi)^2\ $ &
         $-\cos(4\varphi)\sin\bigl(\half\alpha\bigr)^2$\\[1ex]\hline\hline 
  \end{tabular}
\end{table}

After the signal's polarization qubit is converted to a path qubit, we have
the idler-signal wave function
\begin{equation}\label{eq:3}
     \Psi=\frac{1}{\sqrt{2}}\sqcol{\sin(2\varphi)\\ -\cos(2\varphi)}
                 \otimes\column{1\\0}
      +\frac{1}{\sqrt{2}}\sqcol{\sin(2\varphi)\\ \cos(2\varphi)}
                 \otimes\column{0\\1}\,,
\end{equation}
which exhibits the wave functions 
${\sds{\sqcol{v\\h}}=\sds{\sqcol{\sin(2\varphi) \\ \mp\cos(2\varphi)}}}$ 
for the polarization states of the idler corresponding to 
signal path $\sds{\left\{1\atop2\right\}}$ in the interferometer.
These two idler states are told apart by the USD, the three-outcome
measurement that projects on $\idlerwf{3}$, $\idlerwf{4}$, or $\idlerwf{5}$ with
\begin{equation}\label{eq:5}
  \left.\begin{array}{@{}c@{}}
      \idlerwf{3} \\ \idlerwf{4}
    \end{array}\right\}=\frac{1}{\sqrt{2}}\sqcol{\cos(2\theta) \\ \mp1}\,,\quad
  \idlerwf{5}=\sqcol{\sin(2\theta) \\ 0}\,.
\end{equation}
We set the HWP at the USD at angle $\theta$ given by
${\cos(2\theta)=-\cot(2\varphi)}$, for ${22.5^\circ<\varphi<45^\circ}$, and
then have the coincidence probabilities listed in Table~\ref{tbl:coincide} for
the detection of the signal in particle mode, i.e., with the NPBS removed from
the MZI. 
Indeed, upon detecting the idler by D4, we know that the signal surely
took path~1; likewise for D3 and path~2. 
We cannot infer the signal path, not at all, when the idler is detected by D5,
which is the inconclusive outcome of the USD.
Weak path marking is realized by $\sin(2\varphi)\lesssim1$ and
$0\lesssim\cos(2\varphi)\ll1$, i.e., $\varphi\lesssim45^\circ$; then, the
probability for the inconclusive outcome, equal to 
${1-2\cos(2\varphi)^2=-\cos(4\varphi)}$, is close to~$1$.

\begin{figure}
  \centering
  \includegraphics{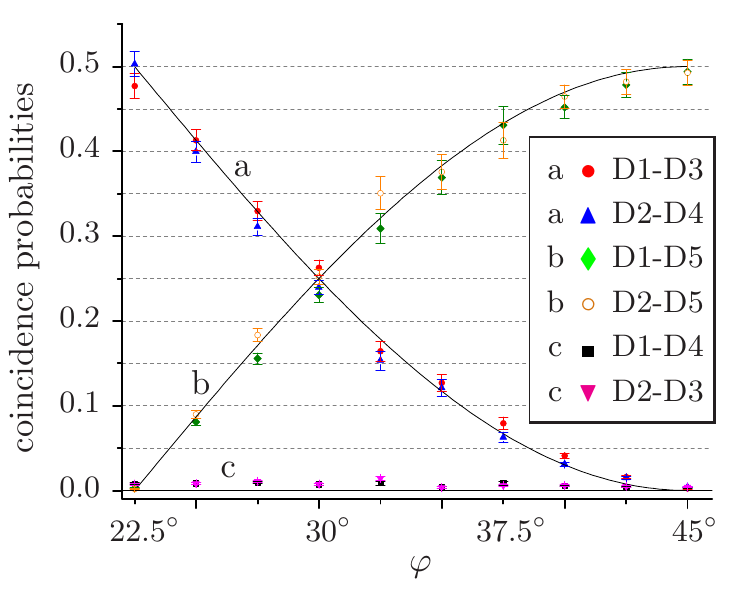}
  \caption{\label{fig:MUDtest}%
   USD outcomes and signal paths. 
   The MZI is operated in particle mode (NPBS removed), so that a signal
   detected by D1 has taken path~2, and likewise for D2 and path~1.
   The solid curves are the theoretical predictions of
   Table~\ref{tbl:coincide} (top part).
   The data confirm in particular that the paths are correctly identified by
   clicks of D3 and D4:  
   There are no D1-D4 and D2-D3 coincidences (curve ``c''). 
   The inconclusive outcomes are also as predicted: 
   The D1-D5 and the D2-D5 coincidences are equally frequent (curve ``b'').
   Here, and also in \Fig[Figs.~]{D5+V}--\Fig[]{balancedMZI}, 
   the error bars indicate one
   standard deviation as obtained from suitable simulations of Poisson
   processes.}
\end{figure}

\heading{Testing the source and the USD}
We test the quality of the two-photon source and the reliability of the USD by 
operating the MZI in particle mode (NPBS removed) so that detection of the
signal by D1 or D2 identifies path~2 or path~1, respectively. 
\Fig[Figure~]{MUDtest} shows the measured coincidence probabilities together
with the theoretical values for an ideal experiment in
Table~\ref{tbl:coincide}.
The agreement between the theory and the experiment is very satisfactory.
In particular, the virtual absence of D1-D4 and D2-D3 coincidences (curve
``c'') confirms that the signal follows path~1 if D4 detects the idler, and
likewise for path~2 and D3.
Further, as predicted, we cannot infer the signal path if the idler is
detected by D5 since the D1-D5 coincidences are as frequent as the D2-D5
coincidences (curve ``b'').
The data from another calibration experiment are reported in the Appendix.

\heading{Sorted subensembles}
The three outcomes of the USD identify three subensembles of signals and a 
corresponding way of blending the mixed signal state with the density matrix
\begin{equation}\label{eq:6}
  \rho=\tr[_{\mathrm{idler}}^{\ }]{\Psi\Psi^{\dagger}}
  =\half\column[cc]{1 & -\cos(4\varphi) \\ -\cos(4\varphi) & 1}
\end{equation}
from three ingredients,
\begin{eqnarray}\label{eq:7}
  \rho&=& \cos(2\varphi)^2\column[cc]{1 & 0 \\ 0 & 0}
         +\cos(2\varphi)^2\column[cc]{0 & 0 \\ 0 & 1}\nonumber\\
        &&\mbox{} +\bigl[-\cos(4\varphi)\bigr]
                   \half\column[cc]{1 & 1 \\ 1 & 1}\,,
\end{eqnarray}
associated with detecting the idler by D4, D3, or D5.
The signals in the first subensemble follow path~1, those in the second
subensemble follow path~2, whereas the path of a signal in the third
subensemble is unknown --- in fact, it is unknowable.

The combined effects of the PS in path~2 and the NPBS on the signal
state yield the transformation
\begin{equation}\label{eq:8}
  \rho\to\half\column[cc]{1-\cos(4\varphi)\cos(\alpha) &
                          \I\cos(4\varphi)\sin(\alpha) \\
                          - \I\cos(4\varphi)\sin(\alpha) &
                          1+\cos(4\varphi)\cos(\alpha)}\,, 
\end{equation}
so that
\begin{eqnarray}\label{eq:9}
  \left.\begin{array}{@{}c@{}}
  \mathrm{prob}(\mathrm{D1}) \\ \mathrm{prob}(\mathrm{D2})  
  \end{array}\right\}&=&\half\bigl[1\mp\cos(4\varphi)\cos(\alpha)\bigr]
\nonumber\\ &=&\half\cos(2\varphi)^2+\half\cos(2\varphi)^2
\nonumber\\&&\mbox{}+[-\cos(4\varphi)]
   {\left\{\begin{array}{@{}l@{}}\cos(\half\alpha)^2 \\ 
                                 \sin(\half\alpha)^2\end{array}\right.}
\end{eqnarray}
are the resulting probabilities for detecting the signal by D1 or D2.
These probabilities exhibit interference fringes with visibility 
${\mathcal{V}=\bigl|\cos(4\varphi)\bigr|}$, which equals the probability of
obtaining the inconclusive result from the USD.
For ${\varphi=45^{\circ}}$ (no path marking) and ${\alpha=0}$ (balanced MZI), 
there is perfect constructive interference at the D1 exit and
perfect destructive interference at the D2 exit.

\heading{Inconclusive USD outcome and fringe visibility}
That the fringe visibility for the signals is equal to the probability of
obtaining the inconclusive USD outcome for the idlers is not coincidental. 
We have a signal in the third subensemble of \Eq{7} if D5 detects the idler,
and this subensemble consists of signals with utterly unknowable paths and
full interference strength.
The other two subensembles are made up of signals with a surely known path,
and such signals do not contribute to the interference.
The wave-mode coincidence probabilities in Table~\ref{tbl:coincide}
and the corresponding ensemble sum in \Eq{9} summarize this matter.
The statistical weight $\bigl[-\cos(4\varphi)\bigr]$ of the third subensemble
must, therefore, be equal to the fringe visibility.

\begin{figure}
  \centering
  \includegraphics{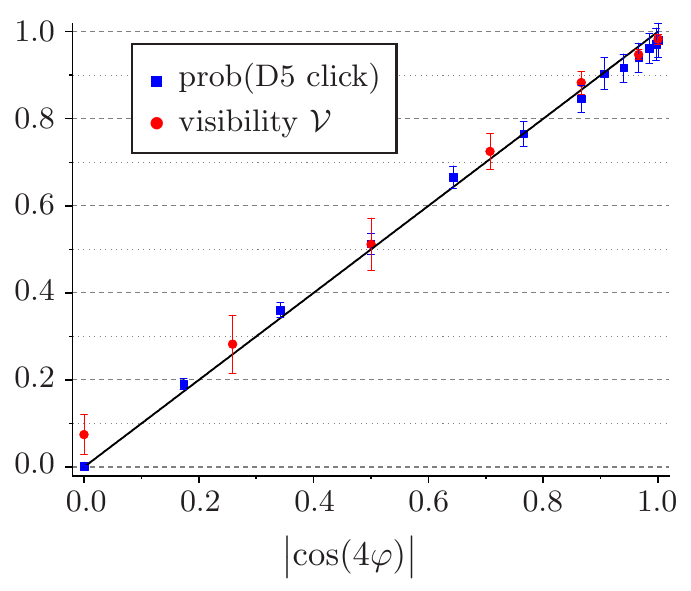}
  \caption{\label{fig:D5+V}%
   Probability of obtaining the inconclusive USD outcome and MZI fringe
   visibility. 
   The experimental data confirm that both quantities are equal to
   $\bigl|\cos(4\varphi)\bigr|$.} 
\end{figure}

These matters are confirmed by the data in \Fig{D5+V}.
Indeed, we have evidence that (i) the probability of getting the inconclusive
USD outcome for the idler (detection by D5) is equal to the fringe visibility 
for the signals (MZI in wave mode), and that (ii) both are equal to
$\bigl|\cos(4\varphi)\bigr|$ in terms of the source parameter $\varphi$.

In passing, we note that the probability of guessing the signal path right on
the basis of the USD outcomes is
\begin{equation}\label{eq:12}
  2\cos(2\varphi)^2+\half\bigl[-\cos(4\varphi)\bigr]=\half(1+\mathcal{K})
\end{equation}
with 
${\mathcal{K}=2\cos(2\varphi)^2=\bigl|\cot(2\varphi)\bigr|\,\mathcal{D}}$. 
In the USD range of ${22.5^{\circ}<\varphi\leq45^{\circ}}$, the path
knowledge ${\mathcal{K}=1-\mathcal{V}}$ 
thus gained is always less than the path distinguishability 
${\mathcal{D}=\sqrt{1-\mathcal{V}^2}}$.
They are equal only for ${\varphi=22.5^{\circ}}$, when ${\mathcal{V}=0}$ and
${\mathcal{K}=\mathcal{D}=1}$.
The purpose of the USD is different, however:
It provides unambiguous knowledge about the signal path as often as possible,
at the price of having no path knowledge at all whenever we obtain the
inconclusive outcome.

\begin{figure}
  \centering
  \includegraphics{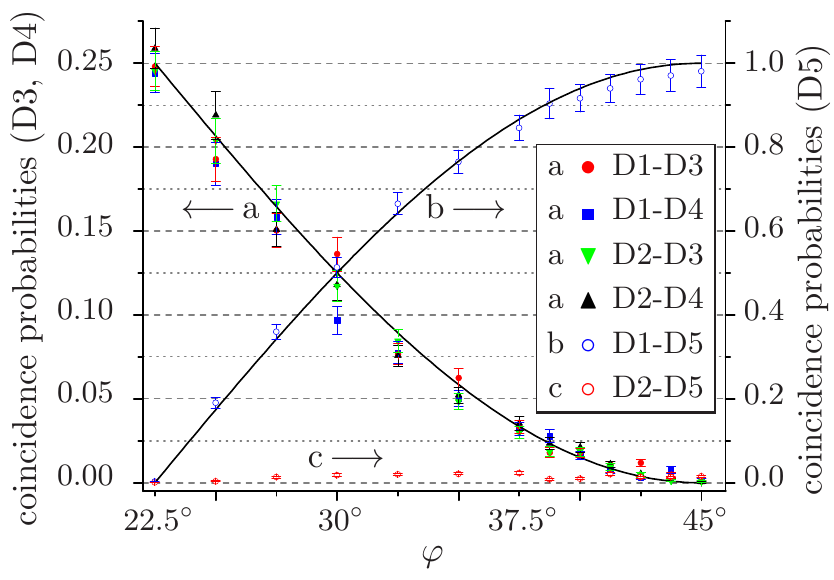}
  \caption{\label{fig:balancedMZI}%
   Coincidence probabilities for the balanced MZI.
   The MZI is operated in wave mode (NPBS in place) and balanced (PS
   set at ${\alpha=0}$).
   The coincidence probabilities for the detector pairs D1-D3, D1-D4, D2-D3,
   and D2-D4 (curve ``a'') are equal and agree well with the theoretical
   prediction in the bottom part of Table~\ref{tbl:coincide} (solid line).
   The $\varphi$ dependence of the D1-D5 coincidence probability is also as
   expected (curve ``b'') and, as predicted, there are no D2-D5 coincidences
   (curve ``c'').} 
\end{figure}

\heading{Balanced MZI with path marking}
Finally, we address the situation encountered in Vaidman's three-path
interferometer 
\cite{Vaidman:13a,Danan+3:13,Zhou+7:17} where the internal MZI is
balanced and photons emerge from the dark output port only because the path
marking disturbs the balance.
So, we set the PS at ${\alpha=0}$ and look for signals detected by D2.
The data in \Fig{balancedMZI} confirm that the paths taken by these signals
are known:
There are D2-D3 and D2-D4 coincidences (path~2 and path~1, respectively; curve
``a'') but no D2-D5 coincidences (curve ``c'').
Whenever there is an inconclusive outcome for the idler (detected by D5), 
the signal is detected by D1 --- at the bright output port (curve ``b''). 
By contrast, the signals with a known path (idler detected by D3 or D4) have
an equal chance of being transmitted or reflected by the NPBS.
This is demonstrated by the equal coincidence probabilities recorded for
D1-D3, D1-D4, D2-D3, and D2-D4 (curve ``a'').
All of these observations are exactly as expected.

\heading{Summary} 
Our experiment realizes a two-path interferometer with unambiguous path
discrimination.
All data are in very good agreement with the theoretical predictions
in Ref.~\cite{Englert+4:17}, which derive from the standard quantum formalism.
We confirm, in particular, that each photon that emerges from the dark output
port of the balanced interferometer has a known path.
This confirmation has an immediate bearing on the experiments by Danan
\textit{et al.} \cite{Danan+3:13} and Zhou \textit{et al.} \cite{Zhou+7:17}, 
inasmuch as it implies that all photons will have a known path through
Vaidman's three-path interferometer \cite{Vaidman:13a} if a suitable
unambiguous path discrimination is performed.

\heading{Acknowledgments}
We are grateful for the insightful discussions with Hui Khoon Ng and the
invaluable advice by Gleb Maslennikov.  
This work is funded by the Singapore Ministry of Education (partly through the
Academic Research Fund Tier 3 MOE2012-T3-1-009) and the National Research
Foundation of Singapore.

\begin{center}
\textbf{APPENDIX}  
\end{center}

\begin{figure}[!b]
  \centering
  \includegraphics{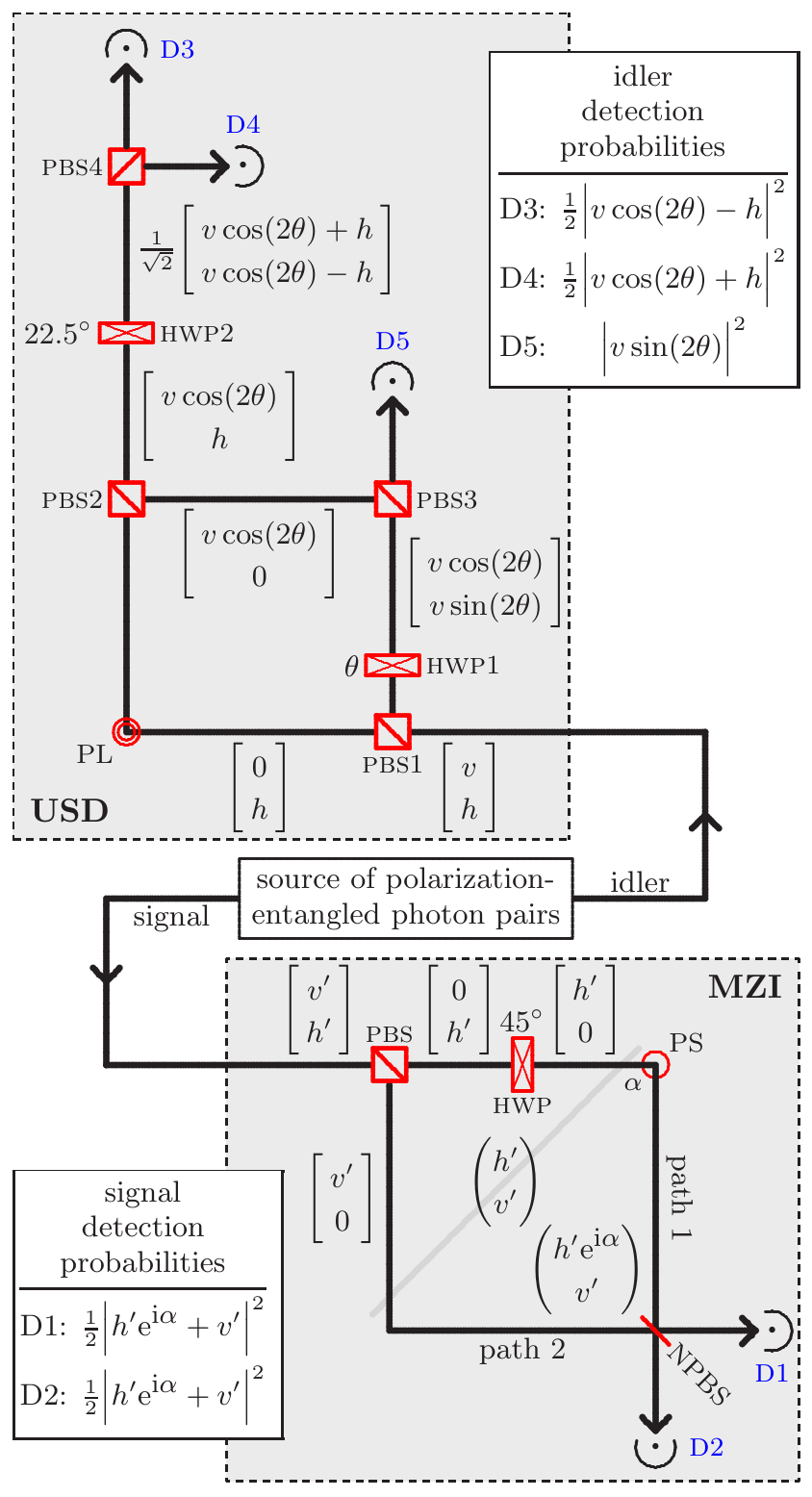}
  \caption{\label{fig:scheme2}
  A more detailed schematic of the experiment in Fig.~\ref{fig:scheme}.
  The polarization states of the signal and idler are indicated at the
  intermediate stages.  
  The insets show the detection probabilities for the various detectors.}
\end{figure}

\heading{Some details of the experiment}
Figure~\ref{fig:scheme2} shows a schematic of the experiment depicted in
Fig.~\ref{fig:scheme}, with a detailed step-by-step account of how the
polarization states of the signal and idler are changed as they traverse the
setup until being finally detected by one of the five detectors.

On the idler side, the polarization qubit with the wave function
$\sds{\sqcol{v\\h}}$ is processed by the USD setup. Eventually, the idler is
detected by D3 or D4 or D5 with the indicated probabilities.

On the signal side, the polarization qubit with the wave function
$\sds[0pt]{\sqcol{v'\\h'}}$ is converted into the path qubit with the wave
function $\sds{\column{h'\\v'}}$, and the signal is eventually detected by D1
or D2. 
The indicated probabilities apply when the MZI is operated in wave mode (with
the NPBS in place); in particle mode (NPBS removed), the detection
probabilities are $\bigl|v'\bigr|^2$ for D1 and $\bigl|h'\bigr|^2$ for D2.

\heading{Source of entangled photon pairs}
The setup for the source of the polarization-entangled photon pairs is 
reported in our earlier work \cite{Dai+4:13}. 
Two beta-borium borate (BBO) crystals of 2\,mm thickness with orthogonal optic
axes are cut at $28.8^\circ$ for non-collinear frequency-degenerate phase
matching. 
The BBOs are pumped by a frequency-stabilized continuous wave diode laser at a
wavelength of 404.6\,nm (Ondax, LM series). 
With non-collinear SPDC, the down-converted photons travel in different
directions with a cone opening angle of $4^\circ$. 
Down-converted photons of narrow spatial bandwidth are collected into
single-mode optical fibers using aspherical lenses with a focal length of
11\,mm, placed at a distance of 900\,mm from the BBOs. 
Using a half-wave plate (HWP) set at an angle $\varphi$ and 
a pair of quarter-wave plates (QWPs) in the pump beam, we convert the pump
polarization from $\sqcol{1\\0}$ to
$\cos(2\varphi)\sqcol{0\\1}+\sin(2\varphi)\sqcol{1\\0}$, where $\sqcol{1\\0}$
and $\sqcol{0\\1}$ represent vertical and horizontal polarizations
respectively. 
The first crystal produces pairs of vertically polarized photons from the
pump's horizontal polarization component, while the second crystal produces
pairs of horizontally polarized photons from the pump's vertical polarization
component, giving us the signal-idler polarization wave function in Eq.(1) in
the main text. 
After leaving the source, and before entering the MZI and USD respectively,
polarization rotations in the optical fibers are compensated for by manual
polarization controllers.

\heading{Photon detection}  
After exiting the MZI and USD, at the various detection ports, the wavelengths
of the signal and the idler are selected by interference filters with a
central wavelength at 810\,nm and a full width at half maximum of 10\,nm. 
The photons are coupled into multi-mode fibers which are fed into
single-photon detectors (Silicon Avalanche Photodiodes, quantum efficiency
of about 50\%, Qutools Twin QuTD). 
The data acquisition system (DAQ) is comprised of a time-to-digital converter
(quTAU, Qutools), which we use to record coincidences of any two detectors
within a coincidence time window of 5\,ns.

\heading{Phase locker and phase shifter}
To implement the phase locker (PL) for the stabilization of the USD, we use an
auxiliary HeNe laser at wavelength 632.8\,nm. 
The locking laser is injected from the unused port of PBS2 in
Fig.~\ref{fig:scheme2}.  
As the optical elements have design wavelength at 810\,nm but not 632.8\,nm,
there are non-zero amplitudes of the HeNe laser propagating through the two
arms of the interferometer in the USD. 
The two beams then interfere at PBS1. At the unused port of the PBS1, we
install a HWP followed by a PBS, and measure the intensity at one of the
output ports with a photodetector. 
The intensity is then converted to a voltage signal, and is fed to a
home-built PID controller. 
The PID controller compares the signal with a fixed reference voltage which
corresponds to the desired phase of the interferometer. 
If the signal deviates from the reference voltage, a feedback voltage is sent
to the the piezo transducer (Thorlabs model TLK-PZT1) attached to the phase
locker mirror. 
The piezo transducer drives the mirror and brings the signal back to the
reference voltage, and hence completes the locking and control of the phase of
the interferometer. 

The implementation of the phase shifter (PS) for the MZI is similar to the one
for USD. 
We inject the HeNe laser from an acute angle reflecting off the interference
filter in front of D1, which then enters the MZI. 
This allows us to lock and control the phase of the MZI using the interference
signal of the HeNe laser at the unused port of the PBS.

\begin{figure}[!t]
  \centering
  \includegraphics{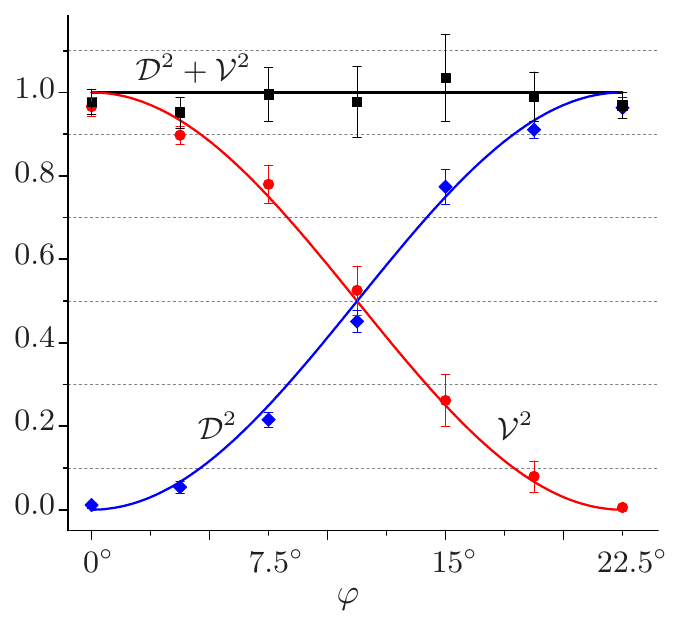}
  \caption{\label{fig:Dsq+Vsq}%
    Path distinguishability and fringe visibility. 
    The plot shows the measured values of the squared path distinguishability,
    $\mathcal{D}^2$, and the squared fringe visibility, $\mathcal{V}^2$, as well
    as their sum.
    The solid lines are the theoretical predictions for an ideal experiment:
    ${\mathcal{D}^2=\sin(4\varphi)^2}$, ${\mathcal{V}^2=\cos(4\varphi)^2}$,
    ${\mathcal{D}^2+\mathcal{V}^2=1}$.
    } 
\end{figure}

\heading{Testing the source and the MZI}
Measurements of the fringe visibility $\mathcal{V}$ and of the path
distinguishability $\mathcal{D}$ for various values of the source parameter
$\varphi$ serve as a test of the MZI, and also as another test of the
photon-pair source. 
The data are reported in \Fig{Dsq+Vsq}.
Again, the comparison with the theoretical values is very satisfactory. 
This is further assurance that our experimental setup has no essential
imperfections. 

Here, the path distinguishability refers to the sorting of the signals
by the outcome of the measurement for error minimization (MEM) on the
idlers~\cite{Helstrom:76}.  
The MEM is an orthogonal measurement that projects on the idler states
${\idlerwf{\pm}=\sds{\frac{1}{\sqrt{2}}\sqcol{1 \\ \pm1}}}$; it is realized by
the setup of \mbox{Fig.~12} in \cite{Englert+4:17} in the place of the
USD in \Fig{scheme}. 
The mixed state of the signals is now blended from two ingredients,
\begin{eqnarray}\label{eq:10}
  \rho&=&\half\rho^{\ }_{+}+\half\rho^{\ }_{-}\nonumber\\
  \text{with}\ 
  \rho^{\ }_{\pm}&=&\half\column[cc]{1\mp\sin(4\varphi) & -\cos(4\varphi) \\
                                -\cos(4\varphi) & 1\pm\sin(4\varphi)}\,.
\end{eqnarray}
Each subensemble has a probability of
$\half\bigl(1+\bigl|\sin(4\varphi)\bigr|\bigr)$ for the more likely signal
path and $\half\bigl(1-\bigl|\sin(4\varphi)\bigr|\bigr)$ for the less likely
path.
We always bet on the more likely path \cite{Englert:96}, and so guess the path
right with a probability~of
\begin{equation}\label{eq:11}
  \half\Bigl(1+\bigl|\sin(4\varphi)\bigr|\Bigr)=\half(1+\mathcal{D})\,.
\end{equation}
This identifies the path distinguishability
${\mathcal{D}=\bigl|\sin(4\varphi)\bigr|}$.
Since we have ${\mathcal{V}=\bigl|\cos(4\varphi)\bigr|}$ for the fringe
visibility, it follows that the duality relation
${\mathcal{D}^2+\mathcal{V}^2\leq1}$ \cite{Jaeger+2:95,Englert:96} is
saturated here.

There are, of course, many more ways of sorting the signals into subensembles.
In particular, there is the quantum erasure sorting in which the subensembles
consist of signals with unknowable paths and exhibit fully visible
interference fringes;
see Ref.~\cite{Englert+1:00} and references therein.


\begin{thebibliography}{10}
\bibitem{fn0}
As we did in \cite{Englert+4:17}, we use columns with round brackets for
path-qubit wave functions and columns with square brackets for
polarization-qubit wave functions.
\end{thebibliography}
\end{document}